\documentstyle[11pt,aaspp4]{article}
\def\etal{{\it et~al.}}
\def\teff{$T_{\rm eff}$}
\def\bele{$^{11}$B}
\def\bten{$^{10}$B}

\def\logg{$\log g$}
\def\gf{{\it gf}}
\def\boi{\ion{B}{1}}
\def\bh{$\log$ N(B)/N(H)}

\def\eps{$ \epsilon $}

\def\lam{$ \lambda $}

\begin{document}

\title{Limits on the Boron Isotopic Ratio in HD 76932}
\author{Luisa Rebull\altaffilmark{1}, Douglas Duncan\altaffilmark{1}, 
Sveneric Johansson\altaffilmark{2}, Julie Thorburn\altaffilmark{1,3},
\& Brian Fields\altaffilmark{4}}

\altaffiltext{1}{Department of Astronomy and Astrophysics, University of
	Chicago (rebull, duncan@oddjob.uchicago.edu)}
\altaffiltext{2}{Department of Physics and Lund Observatory, 
	Lund University, Sweden (SPEK\_SEJ@garbo.lucas.lu.se)}
\altaffiltext{3}{Yerkes Observatory, University of Chicago 
	(thorburn@hale.yerkes.uchicago.edu)}
\altaffiltext{4}{School of Physics and Astronomy, University of Minnesota
	(fields@mnhepw.hep.umn.edu)}

\begin{abstract}
Data in the $\lambda$2090 B region of HD 76932 have been obtained at
high S/N using the HST GHRS echelle at a resolution of 90,000.  This
wavelength region has been previously identified as a likely candidate
for observing the $^{11}$B/$^{10}$B isotopic splitting.  

The observations do not match a calculated line profile extremely well
at any abundance for any isotopic ratio.  If the B abundance previously
determined from observations at $\lambda$2500 is assumed, the
calculated line profile is too weak, indicating a possible blending
line.  Assuming that the absorption at $\lambda$2090 is entirely due
to boron, the best-fit total B abundance is higher than but consistent
with that obtained at $\lambda$2500, and the best-fit isotopic ratio
(\bele/\bten) is in the range $\sim$10:1 to $\sim$4:1.  If the absorption 
is not entirely due to B and there is an unknown blend, the best-fit isotopic
ratio may be closer to 1:1.  Future observations of a similar metal-poor
star known to have unusually low B should allow us to distinguish between
these two possibilities. The constraints that can be placed on the
isotopic ratio based on comparisons with similar observations of
HD~102870 and HD~61421 (Procyon) are also discussed.

\end{abstract}

\section{Introduction}

Light element abundances in stars have much to tell us about big bang
nucleosynthesis (BBN) and galactic chemical evolution.  Only D, He, and
$^7$Li are predicted to be created in standard BBN
(e.g. \markcite{WFH}Wagoner, Fowler, \& Hoyle 1967; \markcite{CST}Copi, 
Schramm, \& Turner 1995); thus, observations of
these elements can provide a check on BBN.  Li, Be, and B (LiBeB)
abundances increase with metallicity, and it is thought that the
observed production is the result primarily of cosmic ray (CR)
spallation.  Since \markcite{mar}Meneguzzi, Audouze, \& Reeves (1971),
it has been thought that the reaction of fast protons and $\alpha$
particles on interstellar CNO is largely responsible for the creation
of $^6$Li, $^7$Li, $^9$Be, \bten, and \bele; our recent work on the
evolution of boron (\markcite{bevol}Duncan \etal\ 1997, 1998) suggests that
this canonical model is incomplete, and that the reverse process of
spallation of heavier nuclei onto protons and $\alpha$ particles may be
more important (see below).  Additionally, simple CR spallation
under-produces B (and Be) at low metallicities and predicts that
\bele/\bten\ $\sim 2.5$, but the observed value in meteorites is
$4.05\pm0.2$ (\markcite{meteor}Chaussidon \& Robert 1995). The
ratio as observed in the interstellar medium (ISM) is
$3.4\pm0.7$ (\markcite{ism}Lambert \etal\ 1998).

Because solving the B isotope problem must happen in the context of
solving the total B evolution with metallicity, it is difficult to
solve the B isotope problem without under- or over-producing the other
light elements. One of the proposed solutions involves a large
low-energy cosmic ray (LECR) component (of fast C,O) which spalls off
protons and $\alpha$ particles in the ISM, and produces \bele\ in
sufficient quantity as to raise the predicted isotopic ratio
(\markcite{rklr}Ramaty \etal\ 1997; \markcite{vcfo}Vangioni-Flam
\etal\ 1996).  \markcite{woos}Woosley \etal\ (1990) have proposed that
neutrino-induced nucleosynthesis in supernovae might produce measurable
amounts of \bele\ but essentially no \bten, so as to increase the
isotopic ratio.  (This method also produces small amounts of $^7$Li, 
as well as certain other species.)
The production by neutrinos solves the boron isotopic
problem, but the $\nu$-process yields are very sensitive to the poorly
constrained neutrino temperatures (and hence spectra); when
incorporating its yields, one must be careful to produce enough
$^{11}$B to change the isotopic ratio to match meteoritic amounts, but
at the same time, not overproduce B to the extent that the observed
B/Be ratio is not fit.  Several proposals have also been made regarding
the destruction of one B isotope but not the other, but these seem
unlikely.  For modeling the evolution of the light elements, the boron
isotopic ratio is the least constrained of all the light element data,
and any additional information will help distinguish between models
with various production and destruction mechanisms.

\markcite{johan}Johansson, Litzen, Kasten, and Kock (1993) measured in
the lab the isotopic splitting of \bele/\bten\ at $\lambda$2089.5 to be
25 m\AA, corresponding to a resolution of about 85,000.  Using the
HST/GHRS echelle (nominal resolution $\sim$90,000), we have observed
the $\lambda$2090 region of HD~76932 ([Fe/H]=$-$1.0).  We will show
that a clear result is not possible at this time.  The data admit both
the possibility that the feature at 2089.5 \AA\ is entirely due to B,
and also the possibility that the feature is a blend of B and an
unknown line.   The resulting best-fit isotopic ratio in the two cases
is different.  The constraints that can be placed on the ratio are
discussed, as well as a future observation which could distinguish
between the two possibilities.

\section{Observations and Data Processing}
\label{sec:obs}

The {\it Hubble Space Telescope} (HST) {\it Goddard High-Resolution
Spectrograph} (GHRS) was used in echelle mode (R$\sim$90,000) to
observe the \boi\ $\lambda$2090 region in HD~76932 on 3 October 1995.
A total exposure time of 5.481 hours was obtained in 9 orbits.  Due to the
spacecraft losing lock on some of the observations, the resultant count
rate in some exposures was low; the final S/N (per point) of the
combined spectrum is $\sim$55.  The total wavelength coverage is
2083.8$-$2094.8 \AA.

The point-spread function (PSF) of the HST/GHRS is comparable to the
width of a diode (1 diode=0.096-0.087 \AA; STScI 1995).  Data was taken
using standard observing strategies to compensate for photocathode
granularity, and to oversample the spectra such that the array of 500
diodes can be used as an array of effectively 2000 pixels.  Thus, each
point is not truly independent; a given diode affects the data recorded
in 4 points.  However, fourteen points define one wing of the
$\lambda$2090 boron line used for this work, and the S/N given above is
per point, not per diode.

The data were processed using standard IRAF/STSDAS processing
algorithms, the most significant of which are POFFSETS, which
determines the relative shifts of the spectra via cross-correlation,
and SPECALIGN, which shifts spectra to align in wavelength space and
then sums them.  The POFFSETS calculations were done in three equal
sections to provide a consistency check; the relative shift separately
determined from the three sections are averaged to calculate a robust
shift for the spectrum as a whole.  Data from different orbits were
combined, weighted by the average number of counts in the raw
spectrum.

The boron line and the region around it were examined in several
subsets of the data.  The feature is quite similar in each data set,
and there are no additional features added by any one set.  An
exception was visit 5, which was significantly noisier than the other
visits and was not included in the final summation.  There were no
unusually large photocathode granularity features that fall near the
line of interest.

\section{Wavelength Accuracies}
\label{sec:waveacc}

The measurement that is sought is very dependent on small 
wavelength shifts.  We have investigated the accuracy limits of the
wavelength scale, both relative and absolute, using calibration
(emission lamp) as well as stellar spectra.  Calibration spectra were
compared with a Pt-Ne line list, and stellar spectra with a Kurucz
model line list.

\subsection{Relative Accuracies}
\label{subsec:relac}

On-board and processing software algorithms compensate for the HST
orbital Doppler shift (corrected for in-flight, $\sim 8$ km/s), and the
Earth's motion around the Sun (corrected for in software, $\sim 30$
km/s).  The FP-SPLIT strategy minimizes the impact of thermal effects,
changes in plate scale ($\lesssim$0.1\%), and geomagnetic effects
($\lesssim$0.5 diode widths).

The limits of wavelength uncertainty intrinsic to the POFFSETS
(cross-correlation) process represent the limits of comparing the data
against itself, independent of any external wavelengths attached to the
data.  The net error measured in this fashion represents the error
expected from photon statistics (as they affect wavelengths), as well
as thermal effects, geomagnetic effects, etc.\ between the
sub-exposures.

The shifts as calculated by the POFFSETS routine for each of three
equal sections in each spectrum were examined.  The mean uncertainty
found this way is 0.274$\pm$0.06 px (=0.0014 \AA).  The uncertainty in
the position of the wavelengths expected from photon statistics alone
is about 1 m\AA, comparable to the net POFFSETS accuracy, so other
effects do not contribute significantly to this source of error.

The emission line profile has been determined by GHRS staff to be
Gaussian with a full-width-at-half-max (FWHM) of $\sim$1.1 diodes.  The
on-board Pt-Ne lamp has 3000 lines measured to $\pm0.2$ m\AA\
(\markcite{reader}Reader \etal\ 1990).  These known lines were used to
calibrate the instrument on the ground, and polynomials used to solve
for the dispersion.  The wavelength scale attached to the data that
comes from STScI is based on that formula, not on any calibrations
taken at the same time as the science data.  For our data, we requested
Pt-Ne calibration exposures at each new carousel setting.

The calibration lamp emission line positions in the vicinity of the
\boi\ $\lambda$2090 feature as measured via Gaussian line-fitting were
compared directly with the line positions as recorded in the Pt-Ne line
list of \markcite{reader}Reader \etal\ (1990).  Over nine lines in all
calibration observations the largest shift between observations is 2.9
m\AA.  Within each observation, the mean differences in positions
average to 0.004$\pm$0.0015 \AA.  This error is a result of errors in
the measurement of the Pt-Ne lines in the lab, errors in the polynomial
fit to the known Pt lines that results in the wavelength file generated
by STScI, and errors in carousel repeatability and wavelength
dispersion.  The overall mean offset of 4 m\AA\ likely represents a
small systematic in carousel position calibration.  The scatter in this
offset, 1.5 m\AA, represents the random error in the carousel position,
and is comparable to the error from POFFSETS.

\subsection{Comparison of model and star lines}
\label{subsec:modellines}

Another way of determining the wavelength scale of the data is to
compare the wavelengths of lines in the stellar data with known
wavelengths; this is an external comparison of wavelength accuracies.
A comparison was made with a best-fit Kurucz synthetic spectrum
computed from a model with stellar parameters as previously determined
from the $\lambda$2500 region (\markcite{bevol}Duncan \etal\ 1997):
[Fe/H]=$-$1.0, \logg=4.0, and \teff=6000 K.  The synthesis is discussed
in more detail in \S\ref{sec:synth} below. The error calculated in this
fashion results from uncertainties in the Kurucz line list (in missing or
unknown blending lines), in the determination of radial velocity, and
in the dispersion/wavelength registration of the data.

Direct cross-correlation of the data and the synthetic spectrum is not
possible (or desireable), as there are too many mismatches between the
spectra.  Therefore comparison was made of all lines between 2083 and
2092 \AA\ which appeared to be unblended.  Line centers of 41 such
lines were measured via Gaussian fits and compared with the line
positions given in the Kurucz list.  The B line was not included since
the reference for the wavelengths was necessarily different than 
that used for the
rest of the Kurucz lines.  The mean offset and standard deviation of
the mean was calculated, and, since there were several lines that were
clearly discrepant between the model and data, outliers that were more
than 1 $\sigma$ away from the mean were dropped, leaving 28 lines.  The
mean offset of these remaining points was $-$1.497$\pm$0.0035 \AA, or
214.9 km/s.  The standard deviation about the mean is 3.5 m\AA, and the
standard deviation of the mean is thus 3.5/$\sqrt{27}$=0.7 m\AA.  We
therefore expect the net error of any given line to be the uncertainty
of the measurement of the line itself (about 3.5 m\AA) and the
uncertainty of the radial velocity determination for the entire
spectrum (0.7 m\AA), added in quadrature, $\approx$ 3.5 m\AA.  Since
the 3.5 m\AA\ uncertainty for the line itself was determined based on
the Kurucz lines, this uncertainty strictly only applies to the other
Kurucz lines.  It is, however, an upper limit for the error on the
\markcite{johan}Johansson \etal\ (1993) measurements of the B lines.
As expected, the absolute wavelength error ($\sim$4 m\AA) is larger
than the relative error ($\sim$1.5 m\AA).

\subsection{Conclusions on wavelength accuracies}
\label{subsec:waveconcl}

The 1.4 m\AA\ limits of POFFSETS represent the accuracy limits of
comparing the data with itself.  This is independent of external
wavelengths attached to the data, and represents the error expected
from photon statistics, thermal effects, geomagnetic effects, etc.  The
error expected from photon statistics alone (in wavelength space) is
about 1 m\AA, comparable to the error determined.

Comparing the wavelength calibration exposures to the laboratory Pt-Ne
line list reveals errors in the polynomial that assigns GHRS
wavelengths and in grating carousel repeatability.  The overall mean
offset is 4 m\AA, and probably represents a systematic error in the
polynomial used for assignment of wavelengths.  Our best estimate of
the random error in carousel placement is 1.5 m\AA.  This error also
represents an internal error, e.g. relative, not absolute, wavelength
errors.

By comparing stellar lines to the lines in the Kurucz model, the
absolute wavelength registration is recovered to an accuracy of at
least 3.5 m\AA.  As expected, the error in the absolute wavelengths is
larger than that for the relative wavelengths.  This error incorporates
uncertainties in the Kurucz line list, in unknown blending lines, in
the determination of radial velocity, and in the dispersion/wavelength
registration of the data.

All sources of error investigated here are significantly less than the
25 m\AA\ isotopic splitting of boron.

\section{PSF Considerations}
\label{sec:psf}

Calibration as well as data files were used to probe the shape of the
point-spread function (PSF).  The PSF of the emission lines in the
wavelength calibrations, although thermally broadened due to the
temperature of the lamp, is still narrower than any stellar line
obtained here.  The mean FWHM of a Gaussian fit to a mean emission line
is $\sim$0.025 \AA.

In the stellar data, the width of the lines is a combination of
instrumental plus stellar broadening.  The stellar broadening has
components from rotation, macro- and microturbulence, and thermal
motion of the atoms. In extremely high S/N data, it is possible to
separately discern the contribution from some of these components, but
in the present case only the net broadening is detectable from the data
alone.  IRAF was used to determine the widths and centers of Gaussians
fit to more than 20 stellar lines.  Many proved to be blended, and in
order to determine the true broadening the narrowest and least blended
lines were selected.  The three narrowest stellar lines all had a FWHM
of 0.0330 \AA\ $=\Delta v = 4.7$ km/s.

\markcite{nissen}Nissen \etal\ (1994), in determining the $^6$Li/$^7$Li
ratio in this star, derived a net FWHM of a radial-tangential profile fit
of 4.7 km/s, exactly the same as the width determined here, despite the
use of a different profile.  The radial-tangential profile is designed
to incorporate radial and tangential motions in the atmosphere; see
\markcite{gray}Gray (1992).  \markcite{nissen}Nissen \etal\ (1994) used
this profile because they detected a slight asymmetry in the red wing
of the Li line, and they determined that this profile was a better fit
than either a pure Gaussian or the standard rotational profile.  Our
data is not high enough S/N for such detailed comparisons.

\begin{figure}[h]
\epsscale{1.0}
\plottwo{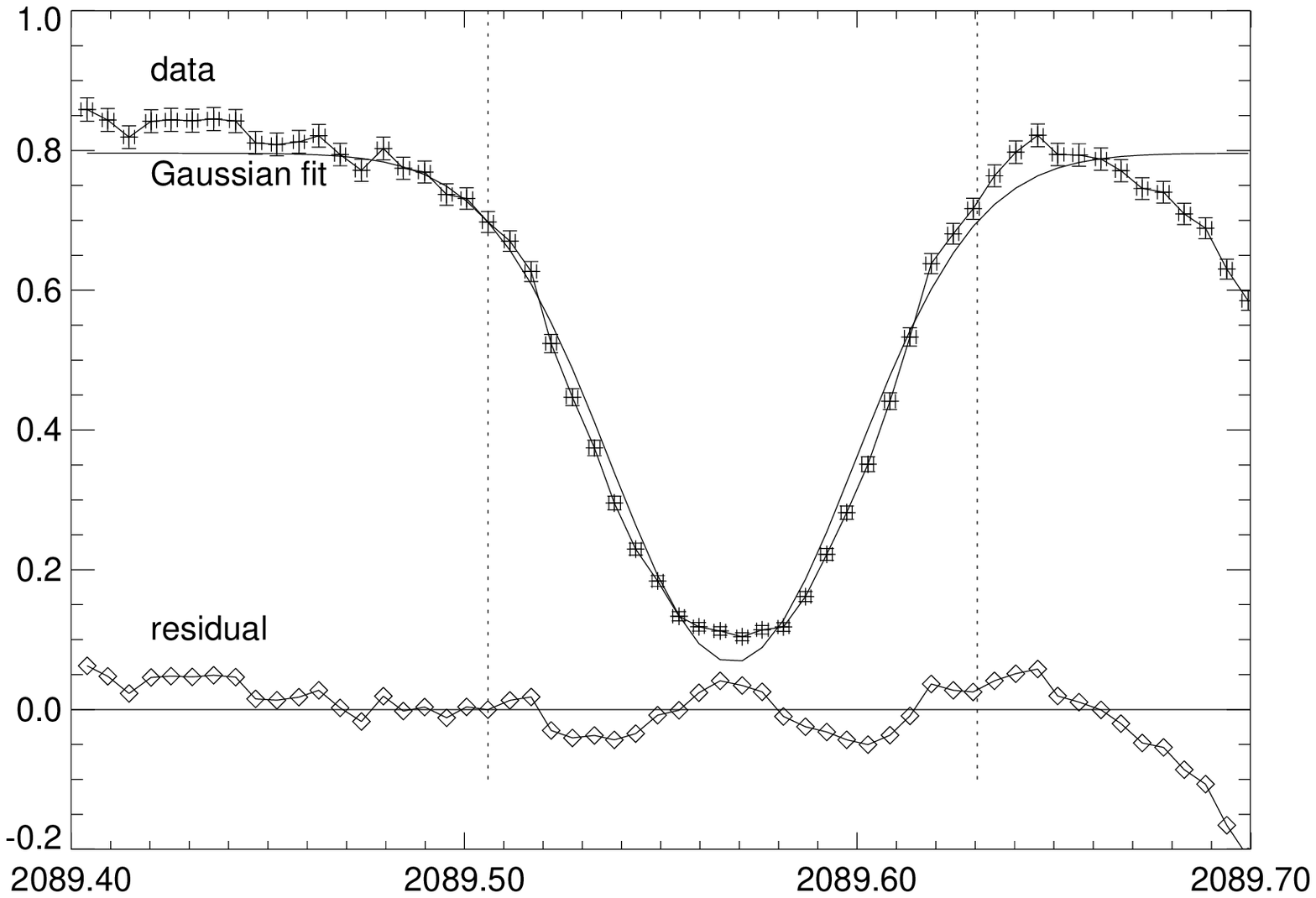}{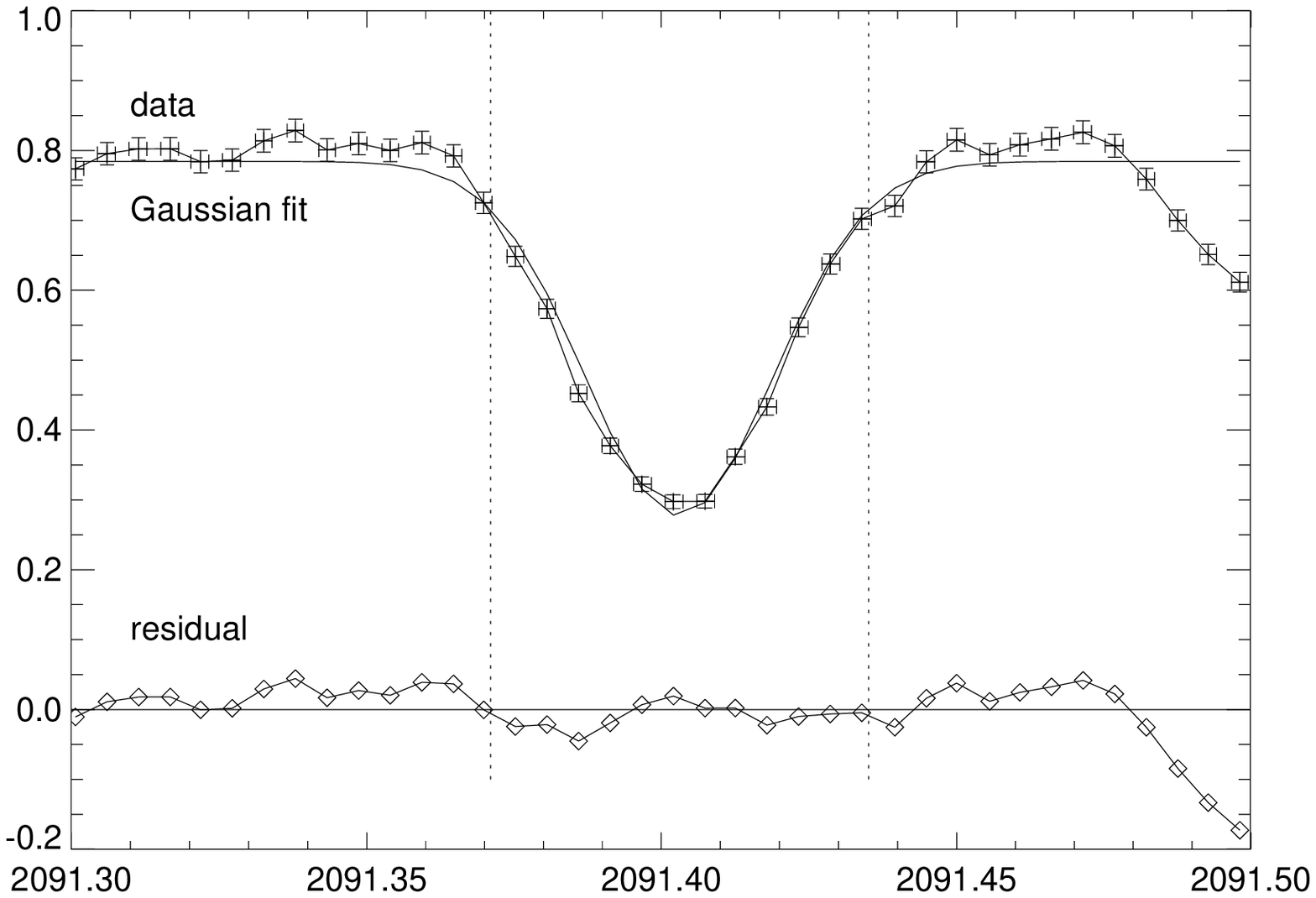}
\caption{Symmetry of the B line (a) and a narrow stellar line at 
$\lambda$2091.4 (b): comparison of a Gaussian to the lines.
Residuals are shown in the lower panel of each Figure.  
Conventions (dotted vertical lines and data point symbols) 
are explained in the text.}
\label{fig:symm}
\end{figure}

For direct comparison to the \boi\ feature, relatively isolated lines
of comparable depth were needed.  The narrowest stellar line found for
this comparison was a line at 2091.4~\AA\ which is not included in the
Kurucz list.  Gaussians were fit to this line and the B feature.  The
width ($\sigma$) in \AA\ of the Gaussian fit to the narrow stellar line
is 0.017, but the width of the boron line is 0.032 (see
Figure~\ref{fig:symm} and further discussion below).  It is clear that
the boron line is much wider than the narrowest stellar lines.  Since
boron is a light nucleus, some additional broadening is expected from
thermal motion of the atoms.  The other identified lines near the B
line are nearly all Fe or Al; B is a factor of about 5 and 2 lighter,
respectively, thus, thermal effects are expected to widen the line by a
factor of about $\sqrt{5}$ and $\sqrt{2}$, assuming that this unknown
line is, in fact, Fe or Al.  However, formal calculations are
preferable, and the Kurucz synthesis incorporates thermal motion
properly.

The boron line as detected here is quite symmetric, as are other lines
in the spectrum.  A simple way of determining the extent of the
broadening and symmetry is to fit a Gaussian to the line and compare
the data to the fit; regardless of the fact that the line is not truly
described by a Gaussian, if it is symmetric, deviations from
Gaussianity should also be symmetric.  Figure~\ref{fig:symm}a compares
a Gaussian fit to the boron line, and Figure~\ref{fig:symm}b does the
same for the stellar line at 2091.4~\AA.  Wings of adjacent features
can be seen on the right.  Residuals are also shown in each Figure.
Conventions introduced in these Figures are used through the rest of
this work.  First, for reference and ease of comparison, (dotted)
vertical lines delineating twice the width of the Gaussian are
superimposed on the Figure as an approximate definition of the region
dominated by the line in question.  Second, data points are indicated
by error bars delimiting photon statistics in the flux direction, and
indicating true wavelength uncertainties (\S\ref{subsec:waveconcl}) in
the wavelength direction.  In this Figure, note that the boron line
deviates from the fit in a symmetric manner, perhaps even more
symmetrically than the 2091.4~\AA\ stellar line.  The red wing of the
boron line is very slightly wider than the blue wing.  It is not clear
whether the symmetry is really symmetry of the stellar lines, or an
artifact of the relatively low S/N of this spectrum.

Extensive tests of Gaussian fits (single and multiple) to these lines
were performed, but none changed the fundamental conclusion that the
boron line is much wider than nearby narrow stellar features.  Since
line profiles are not strictly expected to be Gaussian and especially
since lines in stellar atmospheres do not add as two Gaussians would,
spectral synthesis is needed to make a better attempt at accounting for
the flux in overlapping lines.

\section{Spectral Synthesis}
\label{sec:synth}

Spectral synthesis using the latest Kurucz model atmospheres was done
using the SYNTHE program distributed by \markcite{Kurucz}Kurucz (1993)
on CD-ROM.  This program assumes local thermodynamic equilibrium (LTE)
in determining level populations and calculating the emergent
spectrum.  Scripts written by Steve Allen (UC Santa Cruz) were used to
run the program on Unix SparcStations. The grid of model atmospheres
was that released by \markcite{Kurucz}Kurucz on CD-ROM \#13, which
included in its computation the blanketing of almost 60 million lines,
both atomic and molecular.

\subsection{Boron line parameters}

\markcite{johan}Johansson \etal\ (1993) measured the wavelengths and
\gf\ values of the $\lambda$2090 B feature, and these parameters are
used in the current work and summarized in Table 1.  The line at 2088.9
\AA\ is too blended for abundance analysis; this work is concerned only
with the lines at 2089.5 \AA.  The quoted error on wavelengths is 0.2
m\AA.  The error on the \gf\ values is estimated at 5$-$10\%.

In careful work done previously with line shapes and isotopic
splittings of Li, the hyperfine structure (hfs) of the splittings was
important (\markcite{hobbs}Hobbs \& Thorburn 1994).
\markcite{johan}Johansson \etal\ (1993) state that the isotope and
fine-structure splitting is much larger than the hfs, and therefore hfs
is not a source of error here.  Specifically in the case of Li, one
must worry about where each isotope falls on a curve of growth, as one
component will saturate at a different rate than the other, changing
the shape of the line in ways that might appear to be caused by a
different isotopic ratio.  However, in the present case such errors are
smaller than other errors here and are not considered.

\subsection{Other lines}
\label{subsec:otherlines}

The initial line list used in the spectrum synthesis included the 577
Kurucz lab lines around $\lambda$2090.  There are more than 37~000
predicted lines, but as these predicted wavelengths can be wrong by
several \AA, predicted lines were not used in the detailed analysis.
However, the addition of predicted lines is valid in a statistical
sense, and their inclusion depresses the overall level of continuum
($\sim$0.5\%).  This is less than the other uncertainties in the
normalization process.

Several published line lists were consulted for updates to the Kurucz
line list, but no significant updates were obtained.  Our line list is
available upon request via email to {\tt rebull@oddjob.uchicago.edu}.

We did not adjust the \gf\ values for any of the lines, as we had
insuffient stars to constrain the lines.  Changing the \gf\ 
values for lines close to 2089.5 \AA\ (such as the one seen redward 
of the B line in the Figures) has no effect on the isotopic fits.

\subsection{Previously determined model and initial fits}
\label{subsec:prevmodel}

Our previous work at $\lambda$2500 determined the best-fit Kurucz model
for HD~76932 to have the parameters [Fe/H]=$-$1.0, \logg=4.0,
\teff=6000 (\markcite{bevol}Duncan \etal\ 1997).  Microturbulence, as
in the previous work, was selected from the literature to be 2.0 km/s.
The Kurucz synthesis treats macroturbulence as a Gaussian smoothing
operation, and adds rotational broadening by disk integration.  The net
broadening determined above (\S\ref{sec:psf}) is 4.7 km/s, which was
somewhat arbitrarily apportioned in the synthesis between macroturbulence 
and $v \sin i$, in
the amounts 1.5 km/s for macroturbulence and 4.5 km/s for $v \sin i$.
(i.e. we have used a standard value of 1.5 km/s for macrotubulence,
c.f. \markcite{magain}Magain 1989.)
Our S/N is not high enough to distinguish between these contributions.

The boron abundance reported in \markcite{bevol}Duncan \etal\ (1997)
was 1.82$\pm$0.2 (in the usual notation log \eps(B) = 12.00 $+$\bh).
However, the \gf\ values for the \boi\ $\lambda$2090 feature are known
only from difficult laboratory measurements, and have not been
constrained via observation in a wide range of stars.  We therefore
estimated another 10\% uncertainty due to the $\lambda$2090
\gf\ values.  This brings the effective uncertainty in boron abundance
to $\sim 0.25$ dex.

\begin{figure}[h]
\epsscale{1.0}
\plotone{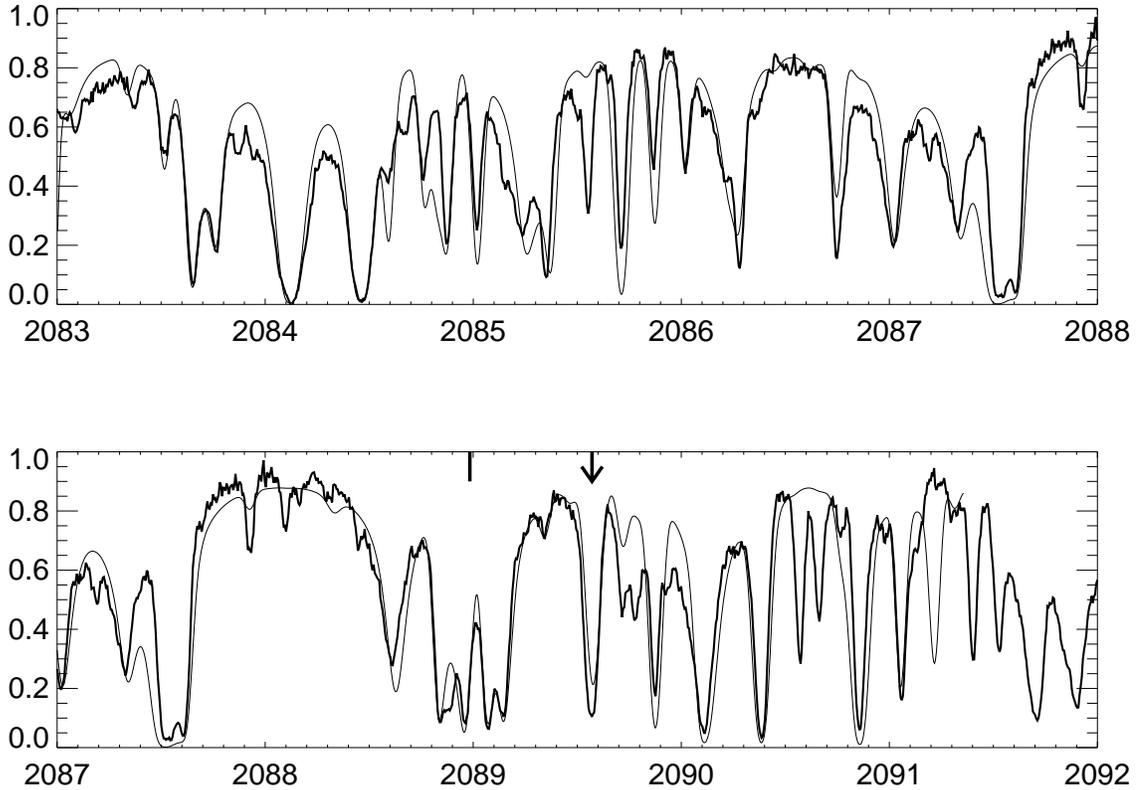}
\caption{Plot of HD 76932 data (thick) and 
Kurucz synthesis (thin). Net broadening is 4.7 km/s, 
total boron abundance is 1.82 dex, and isotope ratio is 1:1.  The 
boron line of interest is marked at 2089.571 \AA.}
\label{fig:kurucz}
\end{figure}

Figure~\ref{fig:kurucz} includes the data and the synthetic spectrum
fit; the net broadening is 4.7 km/s, the boron isotope ratio in the
synthesis is 1:1, and the total boron abundance 1.82 dex.  The boron
lines are indicated at 2089.5 \AA\ (the primary ones in this analysis)
and 2088.9 \AA\ (too blended to be of use here).

It can be seen in Figure~\ref{fig:kurucz} that the boron abundance of
1.82 does not provide a good fit to the B line.  Increasing the boron
abundance to 2.07 dex provides a much better fit.
Note that this new abundance, while higher, is still consistent with
the previous determination at $\lambda$2500, within $\sim 1 \sigma$.

\section{Syntheses with Different Isotopic Ratios} 
\label{sec:ratios}

\begin{figure}[h]
\epsscale{0.5}
\plotone{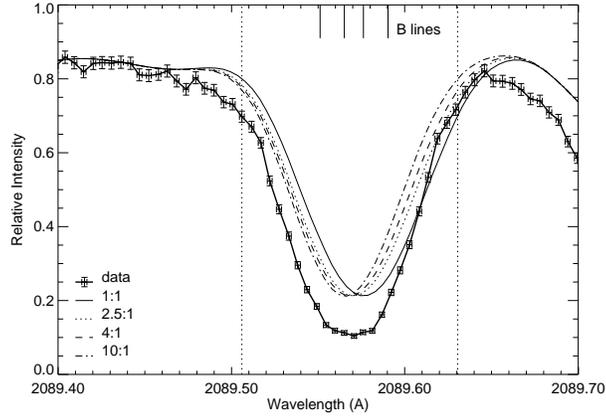}
\caption{Plot of B line and syntheses with total log \eps (B)=1.82, 
at isotopic ratios of 1:1, 2.5:1, 4:1, and 10:1.
The locations of the B lines are indicated; vertical dotted
lines are as in previous figures.}
\label{fig:b182}
\end{figure}

\begin{figure}[h]
\epsscale{1.00}
\plottwo{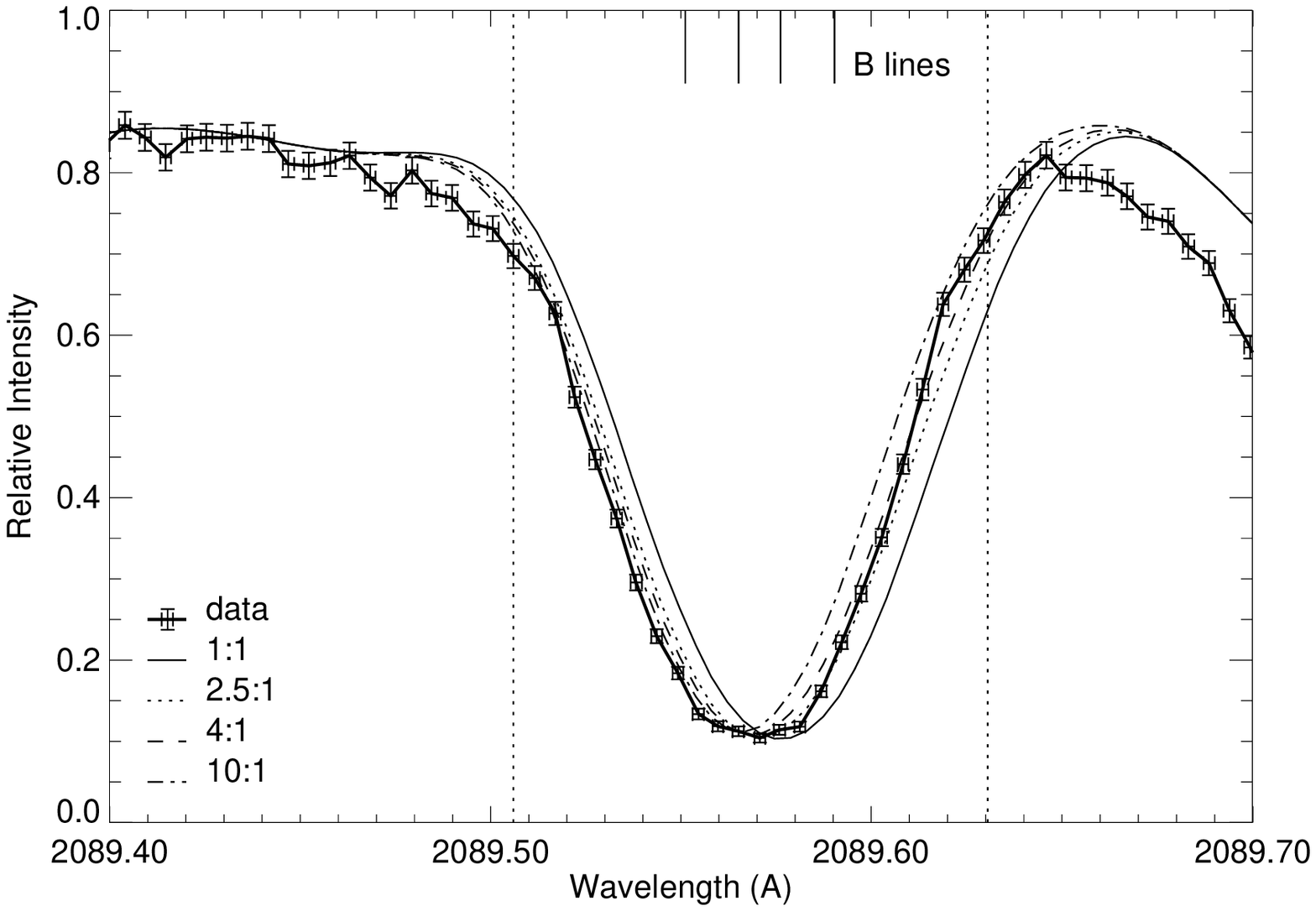}{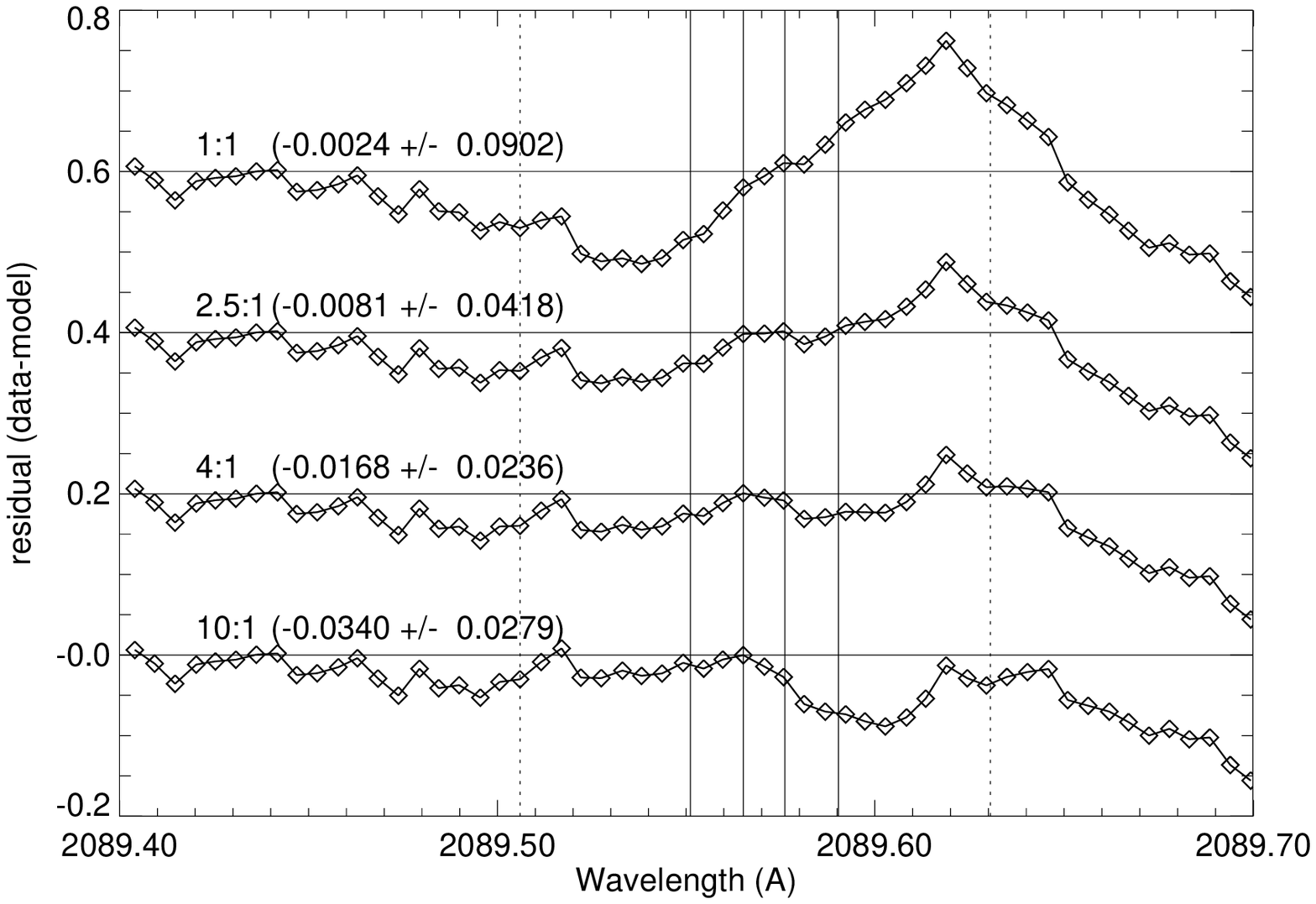}
\caption{B line and syntheses (a) and residuals (b) for total log \eps 
(B)=2.07, at isotopic ratios of 1:1, 2.5:1, 4:1, and 10:1.
Notation is as in previous figures.  Mean and standard deviation of
residuals from the fit are given for each plot in (b).}
\label{fig:b207}
\end{figure}

Models were calculated for isotopic ratios (\bele/\bten) of 1:1, 2.5:1,
4:1, and 10:1, covering the range of galactic chemical evolution model
predictions.  As mentioned in \S\ref{subsec:prevmodel} above, the lower
boron abundance of 1.82 dex, derived from the \boi\ 2500 \AA\ line,
does not provide a good fit to the data.  Consequently, the synthesis
profile does not match the data profile well at any isotopic ratio, as
seen in Figure~\ref{fig:b182}.  Either the boron abundance in the star
is larger, or there is an unidentified line deepening the profile.  The
plot with total log \eps (B)=1.82 is included here for completeness
because of the possibility that the abundance as derived from
$\lambda$2500 is correct and that there is another absorption line
present; see \S\ref{sec:fakeline}.  A synthetic profile that is much
closer to the data can be calculated for a higher B abundance of 2.07
dex, which is within 1$\sigma$ of and therefore consistent with the
$\lambda2500$ determination; this is shown in Figure~\ref{fig:b207}a.
Residuals from the total log \eps (B)=2.07 fit (data$-$model) are
presented in Figure~\ref{fig:b207}b.  For each residual, as in previous
Figures, the dotted vertical lines are for reference and appear at the
same wavelengths in the plots of the line and the residuals.  Within
the region defined by these dotted lines, the mean and standard
deviation of the residuals are indicated on the plot.  The locations of
the boron line components are also indicated with solid vertical lines
between 2089.551 and 2089.590 \AA.

Using this best-fit synthesis of log \eps (B)=2.07, and ignoring the
possibility of a blending line, isotopic ratios of 1:1 and 2.5:1 appear
to be ruled out, as they are too far offset to the red on both wings.
Note that although the data and 10:1 or 4:1 synthesis line shapes agree
relatively well on the blue wing, the shape of the red wing agrees less
well and it is not clear whether the data are more appropriately fit by
the 10:1 or 4:1 profiles.  Deviations from the red wing are smaller for
the 4:1 profile, but the 10:1 fit is better on the blue, and slightly
better overall.  Since the blue wing fits the shape of the profile
reasonably well, it would be expected that the red wing would match to
a comparable degree.  However, in this fit, the red wings of the
synthesis and data are distinctly different.

\subsection{On the shape of the line and possible offsets}

\begin{figure}[h]
\epsscale{1.00}
\plottwo{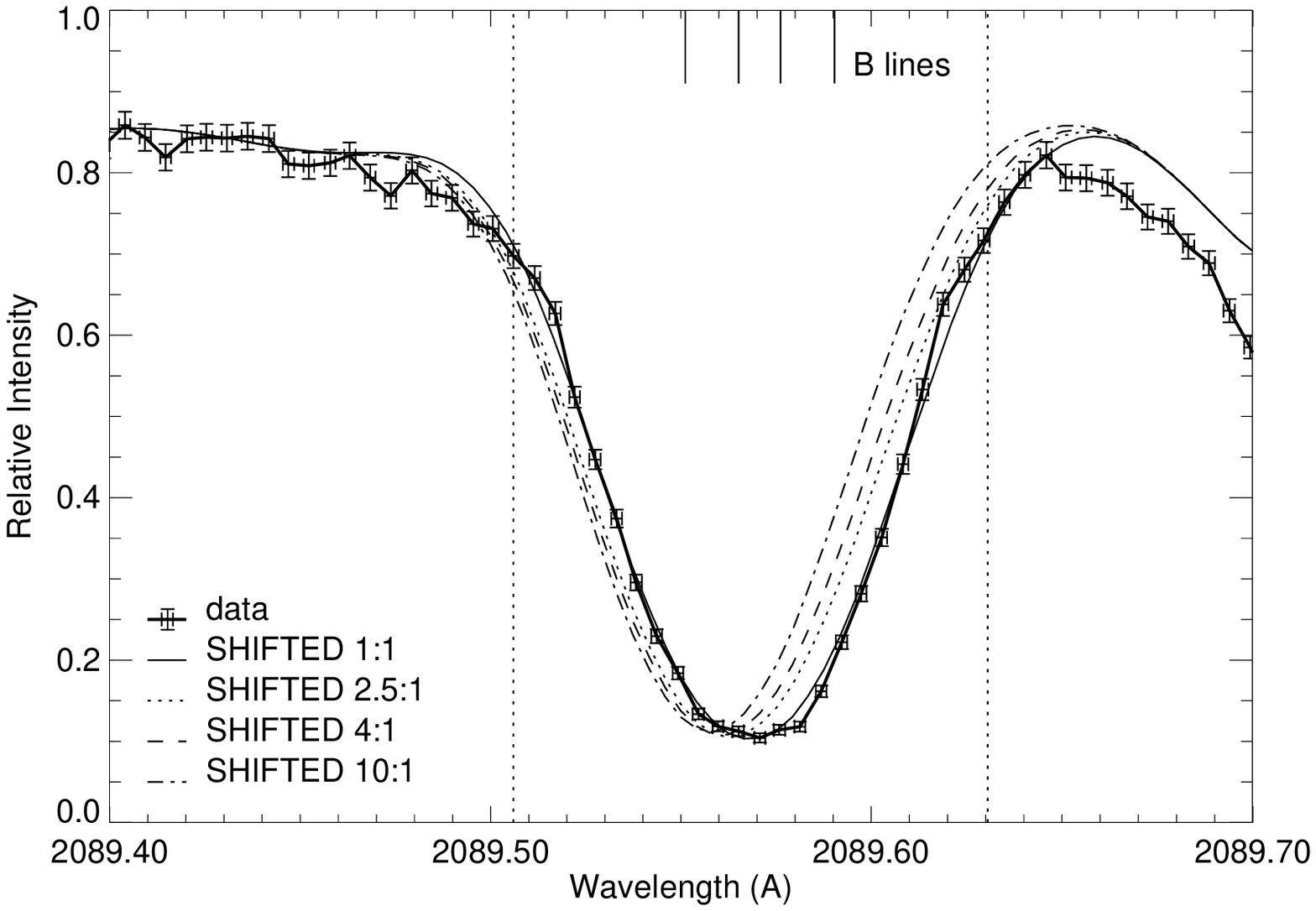}{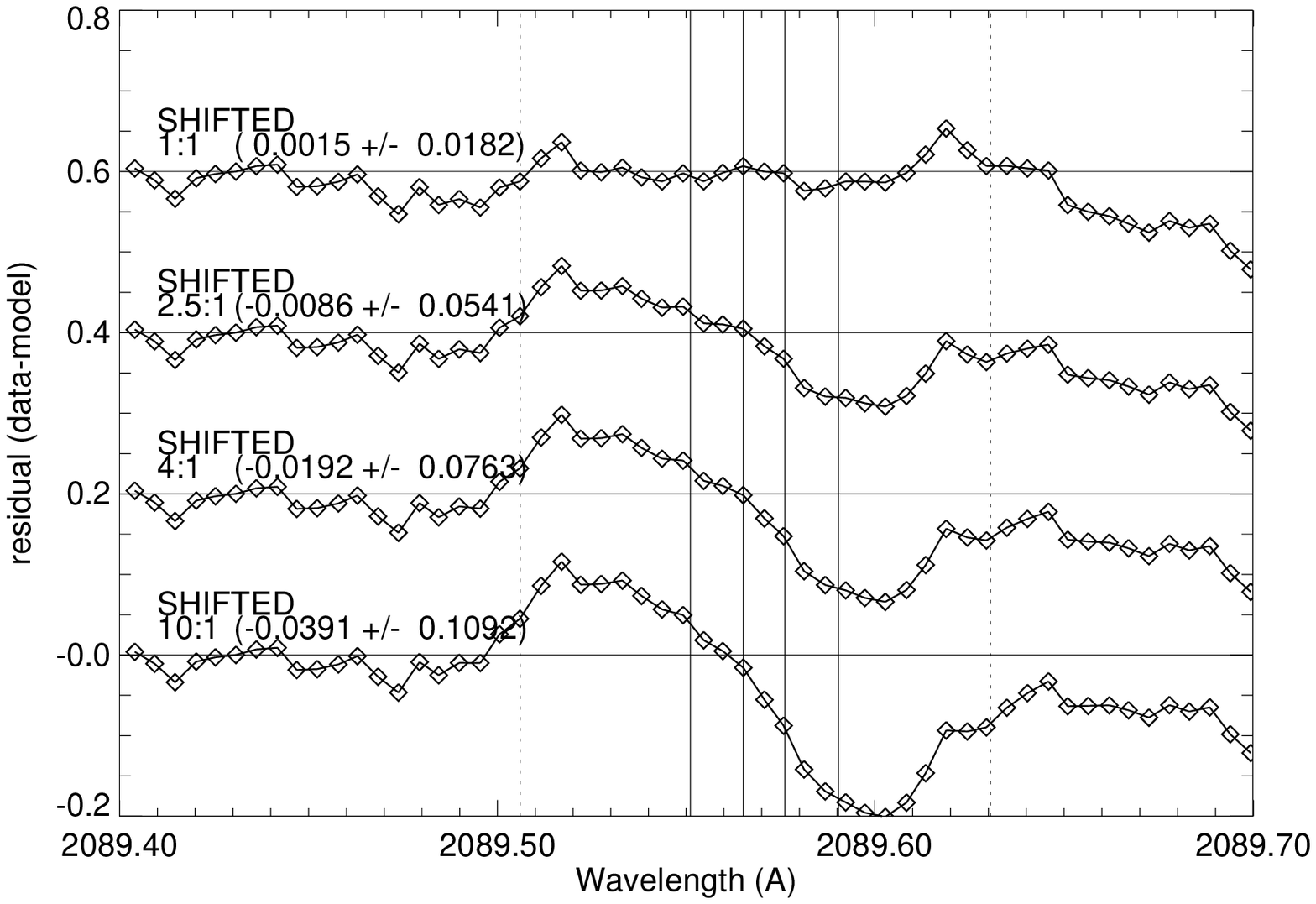}
\caption{Plot of B line and syntheses (a) and residuals (b) for total 
log \eps (B)=2.07, at all four isotopic ratios as before; but with
the syntheses are shifted 0.007 \AA\ to the blue, which does NOT 
provide a good match to the rest of the spectrum. Notation is as 
in previous figures.}
\label{fig:shifted}
\end{figure}

Interestingly, at this higher B abundance, the {\it shape} of the
stellar line is quite well-matched to the synthesis line profile for a
ratio of 1:1, although the {\it position} is not.  In order to force
the 1:1 profile to align with the data, a shift of +0.007 \AA\ is
required in addition to the radial velocity shift of $-$1.497 \AA; see
Figures~\ref{fig:shifted}a and b for such a shifted fit and residuals.
Note that although the overall match with the shifted profile
is quite good, the red wing is still the more discrepant.  A similar
kind of shift does not provide a good match at any other isotopic ratio, 
consistent with the observations in \S\ref{sec:psf} that the
line is very symmetric (other isotopic ratios produce less symmetric
synthetic profiles).

\markcite{johan}Johansson \etal\ quote a wavelength accuracy of 0.2
m\AA\ for the B features, more than an order of magnitude less than 
7 m\AA.  However, since the
wavelengths of the B features are necessarily coming from another
source than the rest of the Kurucz lines, the possibility must at least
be raised that the the boron wavelengths from \markcite{johan}Johansson
\etal\ and the rest of the wavelengths from Kurucz might be offset 
with respect to each other.  However, an offset of as large as 7 m\AA\
seems extremely unlikely.  No source of such an error
was found.

The radial velocity associated with this 7 m\AA\ offset is $\sim 2
\sigma$ away from the best-fit radial velocity as determined in
\S\ref{subsec:modellines} above.  At this new shift the
other lines in the spectrum appear clearly offset with respect to the
synthesis.  Motions of on the order of 1 km/s would be required to
create an offset of 7 m\AA.  It is difficult to imagine a
situation such that the parcel of gas responsible for the B absorption
was moving away from the observer relative to the other lines in order
to account for this offset.  For more discussion on this subject, see
the comparison with HD~102870 (\S\ref{sec:otherstars}) below.

Any systematic offset in \markcite{johan}Johansson \etal\, if present at
\lam2090, might also be present at the \lam2500 B line they also
measured. That spectral region has been more widely studied, and the
results of \markcite{johan}Johansson \etal\ can be roughly compared with other
authors (\markcite{morton}Morton \etal\ 1991; \markcite{edlen}Edl\'en 
\etal\ 1970; \markcite{wagman}Wagman 1937).  
In all cases, all the measurements are quite consistent with
each other, especially since previous work has been done using (nonlinear)
photographic plates, and the most recent work uses a Fourier Transform
Spectometer, estimated at an order of magnitude improved in quality.  
The comparison with other laboratory measurements thus
gives no support to the possiblity of an error in the lab wavelength as
the source of a 7 m\AA\ shift as described above.

\section{Syntheses with Different Isotopic Ratios and a Possible Blending 
Line}
\label{sec:fakeline}

Until this point, we have assumed that the absorption at
$\lambda$2090.5 is largely due to boron.  However, it is possible that
there is an interloping absorption feature from an unknown species
other than boron.  An unknown line, if it is present, could be any of a
wide range of strengths and locations.  As an attempt at quantifying
the impact of a blending absorption feature, a relatively weak line
was selected from elsewhere in the spectrum to use as an artificial
line for several test syntheses.  
The line selected was a neutral silicon line at 2086.747 (log
\gf\ $=-4.275$), and in the synthetic spectrum (with broadening) it
has a core intensity of about 0.4.  This line was added to the line
list at several locations in the boron profile, and the \gf\ value 
adjusted as necessary.

\begin{figure}[h]
\epsscale{1.00}
\plottwo{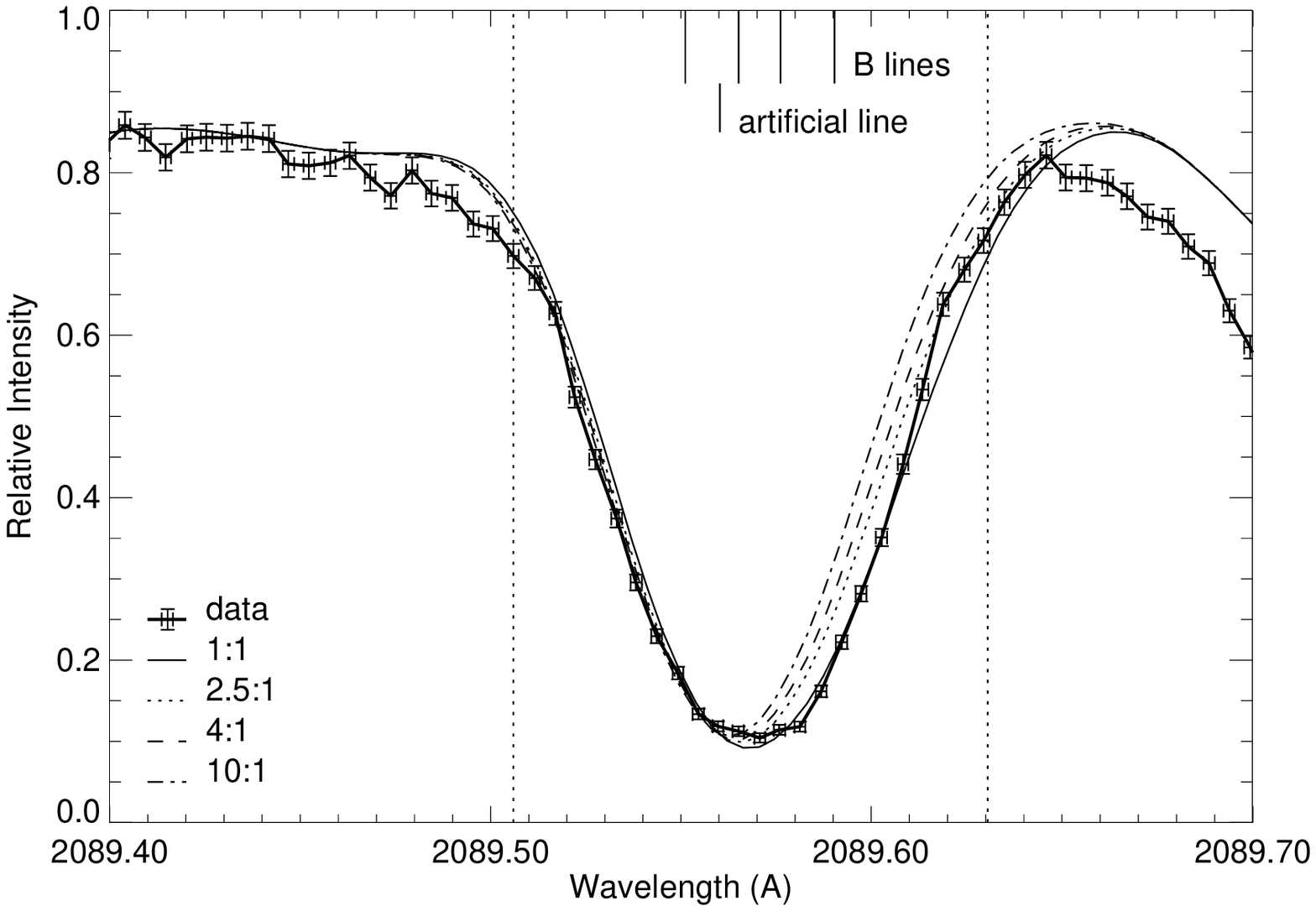}{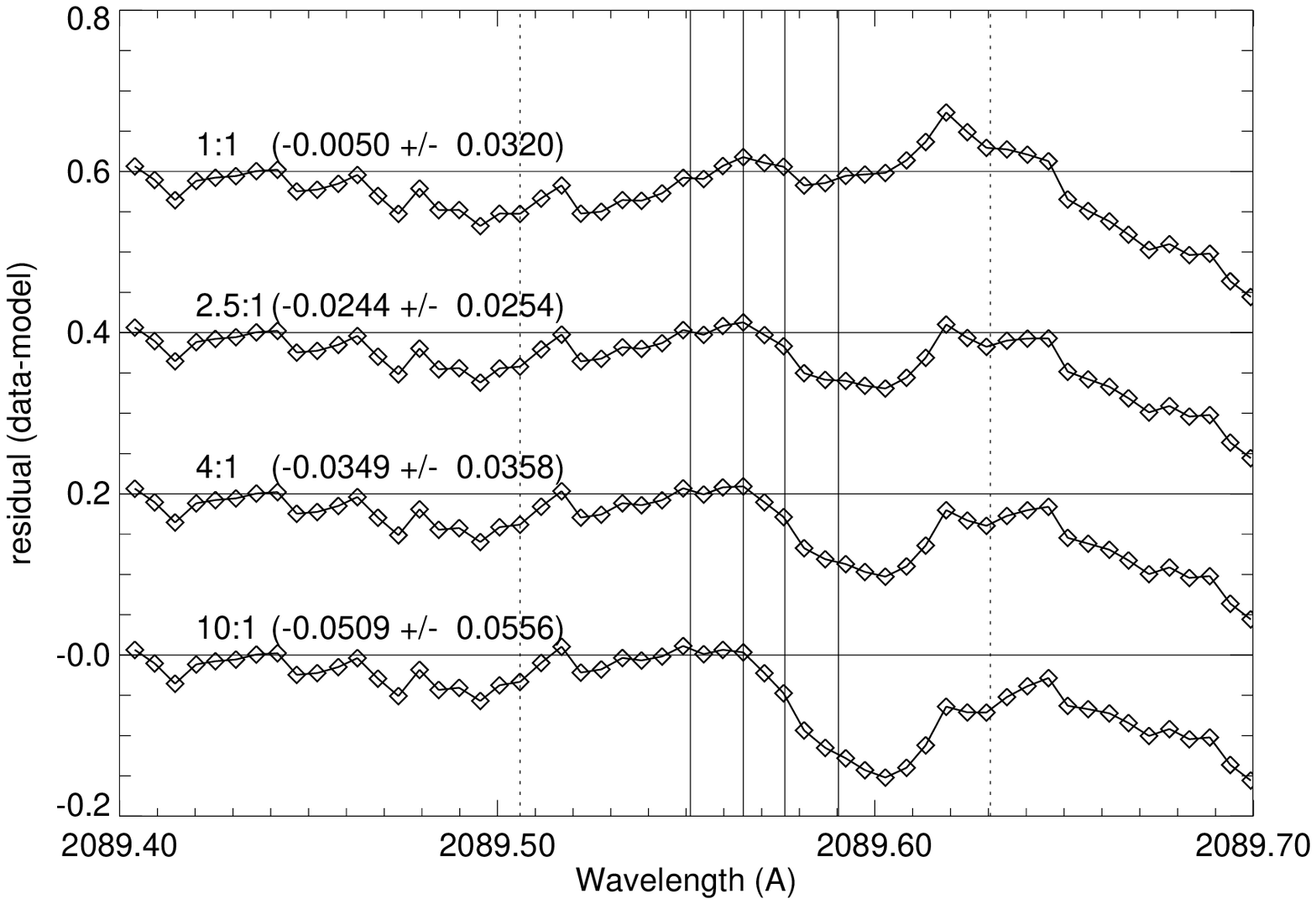}
\caption{Plot of B line and syntheses (a) and residuals (b) for an 
artificial blending line at 2089.560 \AA,
total log \eps (B)=1.82.  Notation is as in previous figures.}
\label{fig:fake182}
\end{figure}

\begin{figure}[h]
\epsscale{1.00}
\plottwo{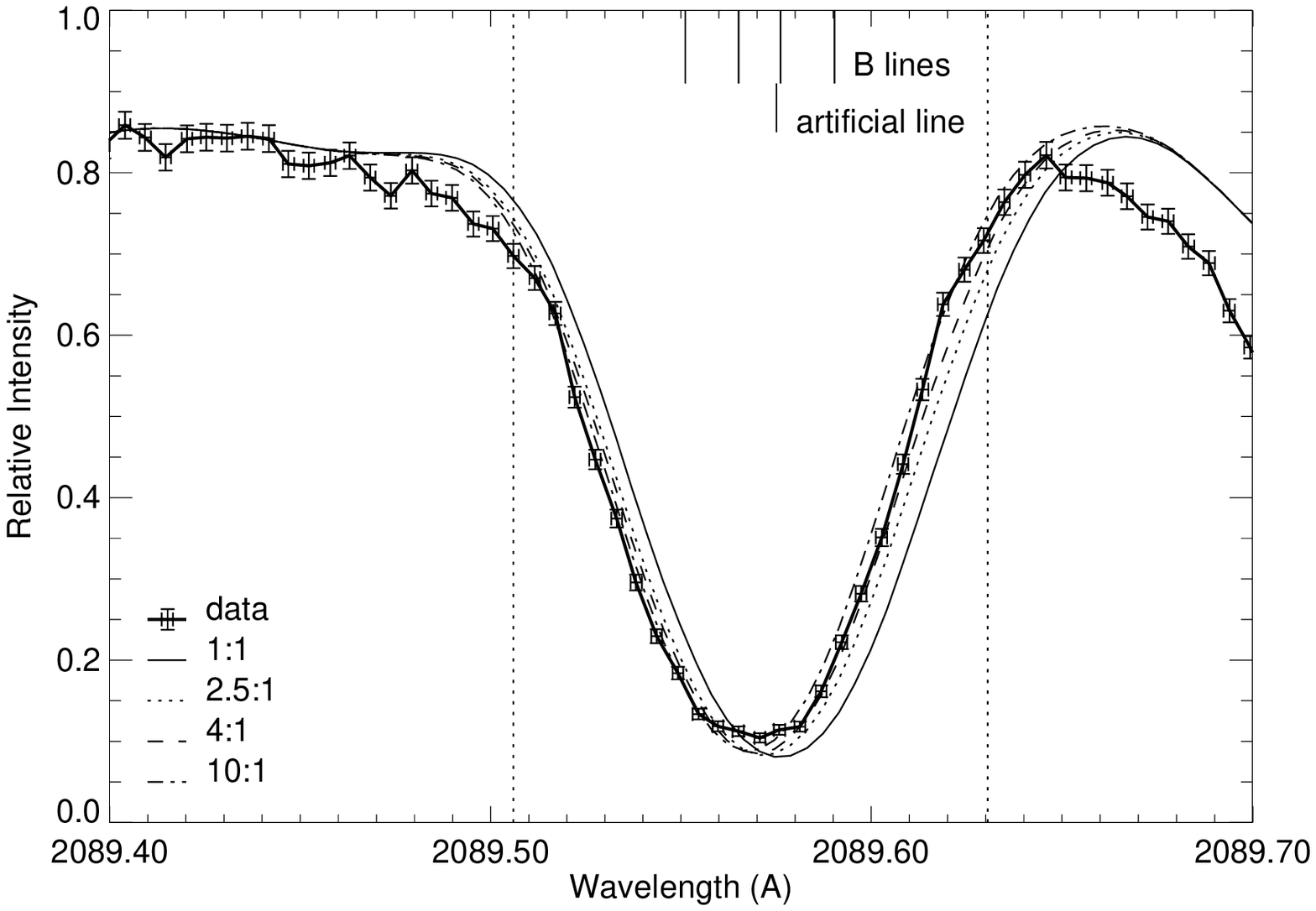}{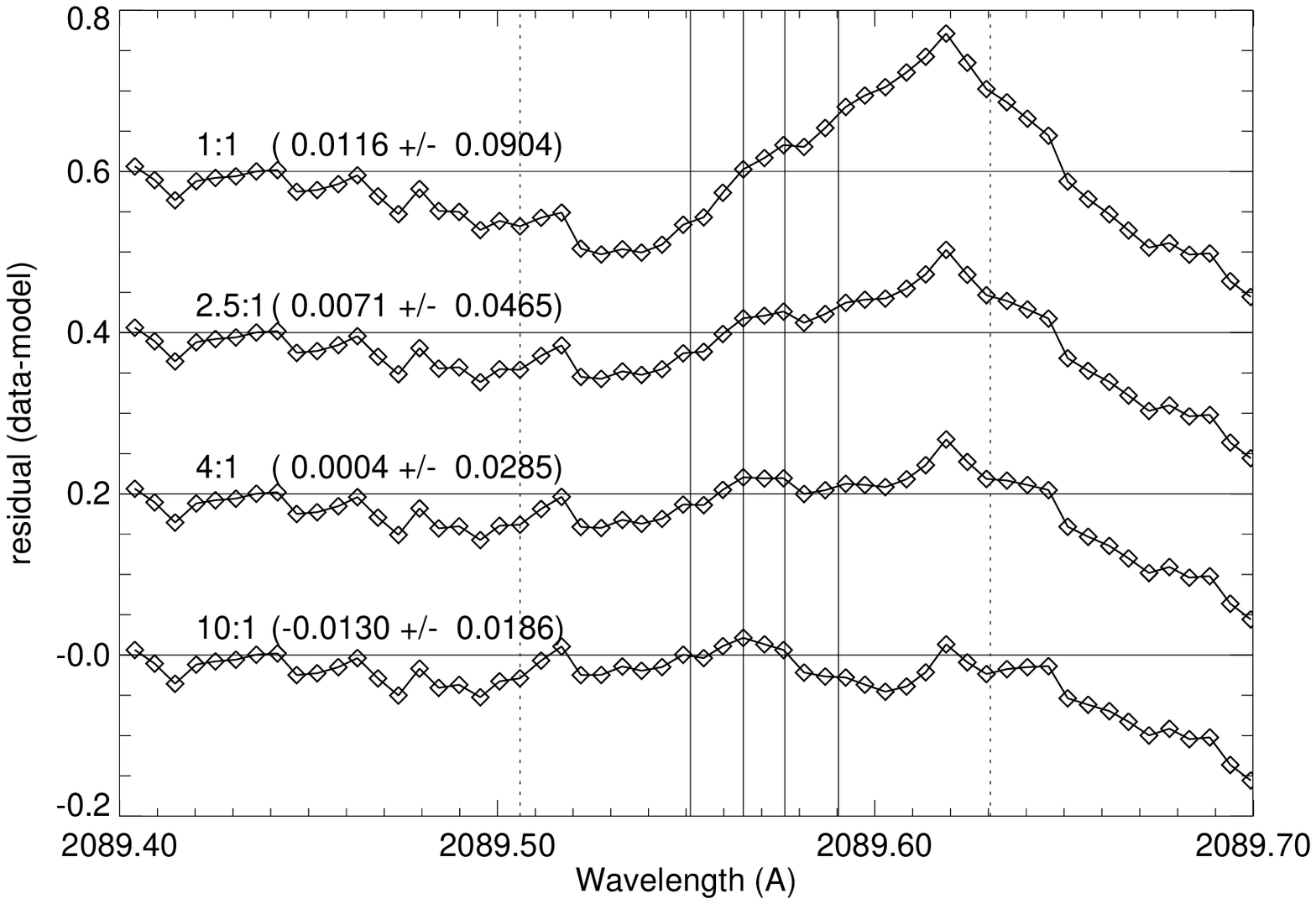}
\caption{Same as Figure 6, but with the 
artificial blending line at 2089.575 \AA, 
total log \eps (B)=2.07.  Notation is as in previous figures.}
\label{fig:fake207}
\end{figure}

Of the parameter space explored, 
the best fits were, for log \eps(B)=1.82, a blending line at 2089.560
\AA, with a \gf\ value of $-$3.875 (shown in Figure~\ref{fig:fake182})
and, for log \eps(B)=2.07, a blending line at 2089.575 \AA, with the
\gf\ value of $-$4.275 (shown in Figure~\ref{fig:fake207}).  As can be
seen in the Figures, in the fits with the higher boron abundance, the
1:1 and 2.5:1 ratios continue to be ruled out, due to
discrepancies on both the blue and the red wings.  The fits with the
lower boron abundance of log \eps(B)=1.82 and the bluer blending line
match the line shape the least well on the red side, less well than the
log \eps(B)=2.07 fits, although all models match the blue side fairly
well.  Unlike all of the other syntheses, however, these favor a 1:1 ratio and
rule out the 10:1 ratio.

In conclusion, if there is a blending line near 2089.575 (or no blend
at all), then the 1:1 and 2.5:1 isotopic ratios are ruled out.  A blend
at about 2089.560, however, allows these ratios.  Making arguments
solely on the shape of the PSF and the consistency with which the
models can duplicate the line shape, it seems likely that any blend, if
it exists, is near 2089.575.  Thus, evidence is in favor of the higher
B isotopic ratio, but a ratio of 1:1 can not be completely ruled out.
Based on these tests in particular, unambiguous conclusions about the
boron isotopic ratio are not possible unless independent information on
possible blends is known.

\section{Additional Sources of Error}

Some of the sources of error usually considered in synthesis fits are
irrelevant here, because the S/N of the data is not extremely high, and
because it is not clear whether or not there are unknown blending lines
present.  Here, the boron line is so deep that it is fortunately not
particularly sensitive to the choice of normalization, and changes in
normalization have no impact on our results.  The microturbulence
cannot be determined from this data, and thus the literature value of
2.0 km/s was used.  The B line is not expected to be particularly
sensitive to errors in \logg, and indeed, large changes in
\logg\ result in little change to the boron abundance (much smaller than
other errors here) and in no change to the best-fit isotopic ratio.

The amount of flux the models predict for this region is sensitive to
the metallicity and \teff\ selected.  A realistic uncertainty on
the metallicity is $\pm$0.1 dex, and the subsequent effect on the boron
abundance is less than 0.1 dex.  In terms of the best-fit isotopic
ratio, a more metal-poor model does not fit any of the profiles well
due to a mismatch on the red wing; for a more metal-rich model, there
is a preference for the 10:1 profile.  A realistic error on \teff\ is
$\pm$75 K; similarly, the boron abundance is changed by less than 0.1
dex.  The hotter model required slightly more boron to fit the data, but did not
result in a change in the best-fit isotope profile.  The cooler model
did not require much change in boron abundance, and fit the 10:1
profile better than the 4:1 profile.

The net error in boron abundance on this determination is at minimum
comparable to that quoted from the $\lambda$2500 determination,
$\pm0.20$ dex.  A good guess for an uncertainty on this result is
$\pm0.25$ dex.  The errors in stellar parameters do not have a
significant effect on the isotopic ratio; in general, if they suggest a
change at all, they are more consistent with the higher isotopic
ratios.

\subsection{NLTE implications}
\label{subsec:nlte}

\markcite{bevol}Duncan \etal\ (1997) used the NLTE corrections of
\markcite{nlte}Kiselman \& Carlsson (1996).  The NLTE correction for
abundances determined at either $\lambda$2500 or $\lambda$2090 
increase the abundance determined using LTE methods.
The $\lambda$2090 correction is 0.35, and that of $\lambda$2500 is
0.17.  Although the LTE determinations of the B abundance in this
star at the two wavelengths are consistent to within the standard
error of measurement, using these NLTE corrections causes the
discrepancy between the abundance
determination to increase.  If the changes suggested by the NLTE 
calculations are correct, then the two abundance determinations
are no longer consistent to within a 1-$\sigma$ error, and such
a difference is thus harder to explain by statistical
fluctuations.  This may be taken as weak evidence that there is an unknown
blending line affecting the $\lambda$2090 feature.  However,
NLTE calculations can place no further constraints on 
the presence of a blending line.

\section{Data for Other Stars: HD 102870 and HD 61421}
\label{sec:otherstars}

We observed HD~102870 as part of this proposal, and 
data exists in the HST archive for HD~61421 (Procyon).  

\begin{figure}[h]
\epsscale{1.0}
\plotone{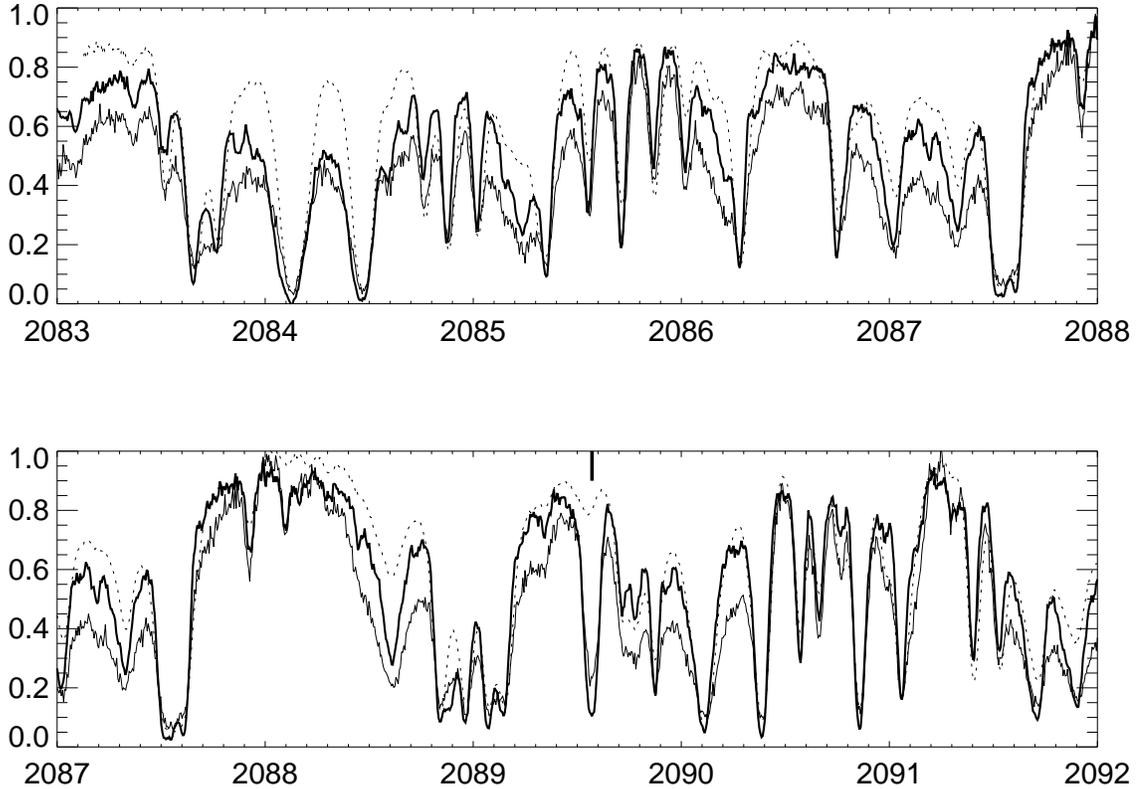}
\caption{Comparison of HD 76932 (thick), HD 102870 (thin), and
HD 61421 (dotted).  The location of the B lines is indicated. 
Note that the data have been normalized to each other, and are
not necessarily properly normalized for comparison to a metal-rich
synthesis.}
\label{fig:76and10and61}
\end{figure}

The stellar parameters of HD~102870 as determined from the literature
are [Fe/H]=+0.2, \logg=4.2, \teff=6100.  The total exposure time for
the HD~102870 data was 59.84 minutes; the resultant S/N per pixel is
$\sim 35$.  The data for HD 61421 (Procyon) are very high S/N, since it
is a very bright star. Total exposure time is 34.28 minutes.
Literature values for Procyon are approximately [Fe/H]=+0.03, \logg
=4.0, and \teff =6500.  In Figure~\ref{fig:76and10and61}, the HD~102870
data are seen as the thin line and HD~61421 is seen as the dotted line.  
Data from both stars were scaled for comparison to the HD~76932
data.  The synthetic spectra for the high-metallicity stars indicates a
pseudo-continuum close to 0.4, rather than the value of $\sim$0.9 seen
in HD~76932.  Many lines which are too weak to affect the HD~76932
synthesis affect these more metal-rich stars, causing the continuum placement
to be very uncertain.  
Thus, direct analysis of the isotopic abundance in these metal-rich 
stars was not attempted.  
Instead, the continuum was scaled to allow the side-by-side comparisons
shown in the figure.

In HD~102870, the B line is at a wavelength extremely consistent with
that in HD~76932, to well within measurement errors at 0.3 m\AA.  There
is no additional relative shift here between the boron line and the
rest of the lines in the spectrum.  If some physical effect is invoked
to explain the 7 m\AA\ shift, it must operate identically in HD~102870,
a much more metal-rich star.  Thus, convective motions do not seem to
be the culprit for creating a 7 m\AA\ offset.

As can be seen in the Figure, there appears to be little boron in
HD~61421.  \markcite{lemke}Lemke, Lambert, and Edvardsson (1993)
studied the $\lambda$2500 region of this star, and concluded that it
was depleted in boron by a factor of at least three.  This star is also
highly depleted in Li and Be, so B depletion may be expected from
stellar structure grounds.  Note that there is little evidence of any
possible blending feature at the wavelength of the B line.  The most
likely candidate for an unknown blending line is an Fe or Al line, but
in this more metal-rich star, such a line should be seen.  It is, of
course, possible that the 500~K difference in temperatures could also
be responsible for the line disappearing.  However, even in the
worst-case scenario of a neutral atomic line, the effects of
metallicity will dominate over the effects of temperature, thus arguing
in favor of no blending line.  Several sample syntheses were run to
confirm that 500~K is not sufficient to completely remove such a
blending line.  The facts that this star has no or little boron and the
$\lambda$2090 line nearly disappears are a strong argument for the
absorption being largely due to B, with no unknown blending line.

Interestingly, the weak absorption line in the Procyon spectrum is
located at 2089.548 \AA, which is distinctly off-center from the B
lines centered at 2089.571 \AA.  Several syntheses with an artificial
line were run in the same vein as \S\ref{sec:fakeline} above, with the
artificial line at the position of this vestigial line.  However, since
this line is distinctly blueward of the B lines (22 m\AA), the changes
in the syntheses are all on the blue wing of the line, not the
more-problematic red side.

Observations of the $\lambda$2090 region in stars closer in
stellar parameters to HD~76932, preferably with known varying amounts
of boron, could clarify the issue of whether or not there is a
blending line.  The B-poor halo stars, measured by Primas \etal\ (1998)
would be ideal targets.

\section{Discussion}

The present work was undertaken to constrain galactic chemical evolution
models.  Since \markcite{RFH70}Reeves, Fowler, \& Hoyle (1970), it
has been known that Galactic cosmic rays (GCRs) are involved in the
formation of light elements.  However, the \bele/\bten\ ratio predicted
by the spallation of cosmic ray (CR) 
protons and $\alpha$-particles off of C, N, and
O in the ISM is 2.5; this is only marginally consistent with the
meteoritic ratio known for some time to be near 4 (most recently
$4.05\pm0.2$, \markcite{meteor}Chaussidon \& Robert 1995), or the ISM
ratio, recently determined to be $3.4\pm0.7$
(\markcite{ism}Lambert \etal\ 1998).  Thus, an additional source of
B is needed.

Three possible contributors to light element abundances in the Galaxy
(besides BBN and AGB stars for Li) have been widely discussed recently.
The mechanism described immediately above, Galactic cosmic 
ray (GCR) protons and $\alpha$-particles spalling off of C, N, and O 
in the ISM, is the mechanism traditionally thought to make
light elements in the Galaxy.
A second process is the neutrino ($\nu$) process
originally suggested by \markcite{woos}Woosley \etal\ (1990) where
$\nu$'s created in the collapsing core of a massive supernova induce
nuclear reactions in the outer metal-rich layers of the star.  Yields
of this process are very uncertain, because the $\nu$ temperature
is poorly constrained by SN models.
The last process most recently discussed is 
low-energy cosmic rays (LECRs), metal-rich material accelerated by
massive stars and spalling off H and He in the ISM.  This is a reverse of
the traditional GCR spallation process. Models of this process 
show interesting results that seem to match the light element evolution;
however,
LECRs, while appearing promising, have an uncertain physical origin.

Vangioni-Flam \etal\ (\markcite{vcfo}1996; hereafter VCFO) developed
models with all three of these components operating, and made specific
predictions for, among other things, the \bele/\bten\ ratio.  The light
elements produced by these processes are listed by isotope in Table 2
(based on a similar table in VCFO). Table 2 also contains rough
predictions for \bele/\bten.  Note that this prediction is often
sensitive to the input parameters of the model, and that, if several
processes are operating, the products must be mixed
together in the ISM, effectively lowering high ratios and raising
low ratios.

The LECR component is constrained by observations by instruments on the
{\it Compton Gamma Ray Observatory}, both detections
(\markcite{bloe}Bloemen \etal\ 1994) and non-detections
(\markcite{murphy}Murphy \etal\ 1996).  The astrophysical source
is uncertain, so VCFO tried several different models and source
compositions, all of which had spectra essentially flat up to a cutoff
energy $\sim$ 30 MeV.  Regardless of variations in the models, the LECR
component as modelled by VCFO seems able to account for adequate Be in
the early Galaxy as well as the approximately linear Be and B evolution
with metallicity seen in the data.
The $\nu$-process, however, only produces selected isotopes, and thus
the yields of this process must be carefully
tuned to maintain agreement with observations of the other light
elements.  Future observations of other $\nu$-process products like 
fluorine might provide additional constraints.  
Since the $\nu$ energy and spectrum are quite uncertain,
this tuning is relatively unconstrained; VCFO tuned the yields to match
the meteoritic \bele/\bten.  The rest of the VCFO models also converge
to the single data point at solar metallicity; the current study is of
course working at a metallicity of $\sim-$1.  If one takes the results
of this work to be that \bele/\bten\ $\sim$4-10, with a slight
preference for the higher value, then this work is consistent with all
of the VCFO predictions.  If the results of this work are taken to be
that the isotopic ratio is low, closer to 1:1, then none of
these specific models are consistent.

\markcite{rklr}Ramaty \etal\ (1997; hereafter RKLR) make 
specific predictions about the \bele/\bten\ ratio.  Working with
the three basic processes listed above, they adjust the energy spectrum
and elemental composition of the LECR using, for example, ejecta specified
by models of massive SNe and WR stars.  They investigate a different 
parameter space than VCFO, taking into account one-step
processes that create LiBeB immediately, as well as two-step processes
that result in LiBeB through an intermediate nucleus, over a wide range
in energies.  
RKLR claim that, since GCR under-produces the isotopic
ratio, and LECR (regardless of specifics) tends to over-produce it, the
two processes operating together provide an easy way to account for the
meteoritic ratio, but the $\nu$ process is still required.  Since the
\bele/\bten\ ratio until now has not been measured in stars, RKLR
predict that \bele/\bten\ will be measured in low-metallicity stars to
be essentially consistent with (and not less than) the meteoritic value
of 4.05$\pm$0.2 (\markcite{meteor}Chaussidon \& Robert 1995); at higher
metallicities, they argue, the Type Ia SNe begin to contribute to the
metallicity (and the cosmic ray flux) but not directly to the B
abundance.  They find that it is difficult for cosmic rays alone to create the
meteoritic \bele/\bten\ abundance, and that either an artificial spectrum is
needed or \bele\ production in (Type II) SNe is required throughout
the life of the Galaxy.  
RKLR make the prediction that if \bele/\bten$>$4, then B/Be
cannot be lower than $\sim$16, and if it is $>$4, then the B/Be and
isotope ratios will be inconsistent.
Similarly based on B/Be
observations, they also predict that \bele/\bten$<$7.  Taking the
results of the present investigation to 
be that \bele/\bten\ $\sim$4-10, with a slight
preference for the higher value, our work is consistent with this
prediction.
Similarly, a lower isotopic ratio of 1:1 is not
consistent with the predictions.

Unfortunately, the relatively wide ratio admitted by present data
do not sharply constrain models. 
Observations of the isotope ratio in more metal-poor stars (assuming
constraints on blending lines obtained from stars similar in parameters
to HD~76932 but B-weak) would better 
distinguish between these models.

\section{Conclusions}

Assuming that the line feature observed at 2090 \AA\ is composed solely
of absorption due to B, then the boron abundance determined here is log
\eps (B)=2.07, which is within 1$\sigma$ of the boron abundance
determined at $\lambda$2500, 1.82.  In this case, low
\bele/\bten\ isotope ratios such as $\sim$1:1 or 2.5:1 are
ruled out, and the data are most consistent with the higher ratios of
$\sim$4:1 or 10:1.

However, the line shape, particularly the red wing, does not match the profiles
from either a ratio of 4:1 or 10:1 particularly well.  It happens to match the
shape of the 1:1 profile (both wings) quite well, but only if an ad-hoc
shift of 7 m\AA\ is introduced.  This shift is an order of magnitude larger
than the expected errors in the Johansson \etal\ laboratory B wavelengths,
and we have not found any other reason for a shift.
It cannot be due to a relative radial velocity
shift between B and the other stellar lines, unless one arbitrarily
introduces a shift $\sim 2 \sigma$ away from the best-fit radial velocity
determined from the lines in our spectrum other than B.
Reasonable changes in stellar parameters are not able to create a
better fit, although some selected parameter changes do favorably
affect the profile shape.  All changes investigated here suggest that
the higher isotope ratios are correct; none suggest that the lower
isotope ratios are correct.  A simple examination of the NLTE
corrections weakly suggest the presence of a blending line.

A blending line might simultaneously explain why the 2090 \AA\ B
feature gives a higher abundance than the $\lambda$2500 lines and why the
profile-fitting was not more sucessful.  In this part of the uv spectrum
many stellar lines remain unidentified.  Assuming that
the true B abundance of this star is really 1.82, we have shown
that if there was a blending line of the right strength near 2089.560
\AA, the isotopic ratio could be low.  
If the B abundance is
really higher, closer to the 2.07 determined here, a blending line
could still be present at 2089.575 \AA.  This blending line does not change
the derived isotopic ratio, but rather adjusts the red wing shape at
the expense of the blue to be more consistent with the observed
profile.

Data from this line region of Procyon (HD 61421) were retrieved from
the HST archive in order to search for a possible blending line.  
Only a vestigial line is observed in Procyon (which is B-weak) at
$\lambda$2090.  With only a 500~K temperature difference between these
stars, it seems unlikely that a significant blend would be present in
HD~76932 but absent in Procyon.  Moreover, the central wavelength of
the vestigial feature in Procyon is offset from the expected B
wavelength by $\sim$20 m\AA, a large difference.  Based on this
comparison, we conclude that there is probably no interloping feature
that disrupts the B feature in the cooler or more metal-poor stars.  We
also conclude that this feature in Procyon is probably not B.
This conclusion could be made much stronger if a star without B and
closer in temperature and metallicity to HD~76932 could be observed.
 
A \bele/\bten\ ratio of 4-10 is consistent with essentially all of the
theoretical predictions made by VCFO and RLKR, which incorporate mixes
of products of 3 processes, GCR spallation, $\nu$-process, and LECRs.
It is inconsistent with a pure GCR origin of B.
If the slight preference for the higher isotopic ratios proved to be 
real, it would give evidence that the $\nu$ process is operating  to 
contribute \bele.  On the other hand, if a blending line is present,
and \bele/\bten\ were as low as $\sim 1$, this would
be inconsistent with pure GCR, and the predictions of
VCFO and RLKR, and thus would challenge the current theoretical picture.
Observations of
one of the recently discovered B-poor halo stars (Primas \etal\ 1998)
would decide between the analyses with and without a
blending line presented here, and a more definitive test of the
theoretical models.

\acknowledgements
We wish to thank Lew Hobbs for many helpful discussions.
We remember our friend and colleague David Schramm,
and acknowledge his support and his comments on the initial manuscript.
This research was based on observations obtained with the NASA/ESA {\it
Hubble Space Telescope} through the Space Telescope Science Institute,
which is operated by the Association of Universities for Research in
Astronomy, Inc., under NASA contract NAS5-26555. 
This research has
made use of NASA's Astrophysics Data System Abstract Service.

\begin{table}[h]
\caption{Boron line parameters, including log \gf\ values for
varying isotopic ratios, from Johansson \etal\ (1993).}
\begin{flushleft}
\begin{tabular}{ccccccc} \hline
isotope & $\lambda$ & \gf\ & log \gf\ for 1:1 & log \gf\ for 2.5:1 &
                log \gf\ for 4:1 & log \gf\ for 10:1 \\
\hline
10 & 208.95898 & 0.170 & $-$1.0706 & $-$1.167 & $-$1.372 & $-$1.770 \\
11 & 208.95650 &       & $-$1.0706 & $-$0.991 & $-$0.894 & $-$0.815 \\
10 & 208.89084 & 0.095 & $-$1.3233 & $-$1.420 & $-$1.624 & $-$2.022 \\
11 & 208.88835 &       & $-$1.3233 & $-$1.244 & $-$1.147 & $-$1.068 \\
10 & 208.95760 & 0.019 & $-$2.0223 & $-$2.119 & $-$2.323 & $-$2.721 \\
11 & 208.95510 &       & $-$2.0233 & $-$1.943 & $-$1.846 & $-$1.767 \\
\hline
\end{tabular}
\end{flushleft}
\end{table}

\begin{table}[h]
\caption{LiBeB production (based on VCFO).}
\begin{flushleft}
\begin{tabular}{ccccccl} \hline
source & $^6$Li & $^7$Li & $^9$Be & \bten\ & \bele\ & \bele/\bten\ \\
\hline
GCR  & yes & yes & yes & yes & yes & 2.5 \\
$\nu$ process & no & yes & no & no & yes & large, if full yields are used.\\
LECR & yes & yes & yes & yes & yes & ranges, 3-5 \\
\hline
\end{tabular}
\end{flushleft}
\end{table}

\end{document}